\input ./style/arxiv-general.cfg
\documentclass[aoas,MSNbibl,nameyear,seceqn,rotating,dvips]{arximspdf}
\makeatletter
   \@ifpackageloaded{graphicx}{}{\usepackage{graphicx}}
\makeatother
\usepackage{dcolumn}
\usepackage{graphicx}
\doi{10.1214/14-AOAS804}% Updated by VTEXPTS2LaTeX.exe, 27.01.2015 07:44
\volume{9}
\issue{1}
\pubyear{2015}
\firstpage{452}
\lastpage{473}
\docsubty{FLA}

\makeatletter
\newcolumntype{d}[1]{D{.}{.}{#1}}

\newcommand{\eqref}[1]{(\ref{#1})}
\newcommand{\dif}{\mathrm{d}}
\makeatother

\begin{document}
\begin{frontmatter}

\title{A two-step approach to model precipitation extremes in
California based on max-stable and marginal point processes\thanksref{T1}}
\runtitle{Two-step approach for spatial extremes}
\thankstext{T1}{Supported by contracts with Environment
Canada and a grant from the University of Connecticut Research Foundation.}

\begin{aug}
% Corresponding author: Jun Yan - jun.yan@uconn.edu% Updated by
%VTEXPTS2LaTeX.exe, 27.01.2015 07:44
\author[A]{\fnms{Hongwei}~\snm{Shang}\ead[label=e1]{hongwei.shang@hp.com}},
\author[B]{\fnms{Jun}~\snm{Yan}\corref{}\ead[label=e2]{jun.yan@uconn.edu}}
\and
\author[C]{\fnms{Xuebin}~\snm{Zhang}\ead[label=e3]{Xuebin.Zhang@ec.gc.ca}}
\runauthor{H. Shang, J. Yan and X. Zhang}
\affiliation{Hewlett Packard Labs,
University of Connecticut and Environment Canada}
\address[A]{H. Shang\\
Hewlett Packard Labs\\
1501 Page Mill Rd\\
Palo Alto, California 94304\\
USA\\
\printead{e1}}
\address[B]{J. Yan\\
Department of Statistics\\
University of Connecticut\\
215 Glenbrook Rd. Unit 4120\\
Storrs, Connecticut 06269\\
USA\\
\printead{e2}}
% \phantom{E-mail:\ }\printead*{e2}}

\address[C]{X. Zhang\\
Environment Canada\\
Climate Data and Analysis\\
4905 Dufferin Street\\
Downsview, Ontario M5H 5T4\\
Canada\\
\printead{e3}}
% \printead{u1}}
\end{aug}

% HISTORY:
%
\received{\smonth{12} \syear{2013}}% Updated by VTEXPTS2LaTeX.exe,
%27.01.2015 07:44
%
\revised{\smonth{12} \syear{2014}}% Updated by VTEXPTS2LaTeX.exe,
%27.01.2015 07:44

% ABSTRACT
%
\begin{abstract}
In modeling spatial extremes, the dependence structure is classically
inferred by assuming that block maxima derive from max-stable processes.
Weather stations provide daily records rather than just block maxima.
The point process approach for univariate extreme value
analysis, which uses more historical data and is preferred by
some practitioners, does not adapt easily to the spatial setting.
We propose a two-step approach with a composite likelihood that
utilizes site-wise daily records in addition to block maxima.
The procedure separates the estimation of marginal
parameters and dependence parameters into two steps.
The first step estimates the marginal parameters with an independence
likelihood from the point process approach using daily records.
Given the marginal parameter estimates, the second step estimates the
dependence parameters with a pairwise likelihood using block maxima.
In a simulation study, the two-step approach was found to be more
efficient than the pairwise likelihood approach using only block maxima.
The method was applied to study the effect of El~Ni\~no-Southern
Oscillation on extreme precipitation in California with maximum
daily winter precipitation from 35 sites over 55 years.
Using site-specific generalized extreme value models,
the two-step approach led to more sites detected with the El~Ni\~no
effect, narrower confidence intervals for return levels and tighter
confidence regions for risk measures of jointly defined events.
\end{abstract}

% KEYWORDS
\begin{keyword}
\kwd{Composite likelihood}
\kwd{estimating function}
\kwd{extreme value analysis}
\kwd{risk analysis}
\kwd{spatial dependence}
\end{keyword}
\end{frontmatter}

%s1 #&#
\section{Introduction}
\label{sec:intro}

Environmental extreme data are often spatial in nature, as data
are recorded at a network of monitoring stations over time.
Extreme weather and climate events may also exhibit spatial
dependence because their occurrences are influenced by
atmospheric circulation of a very large spatial scale.
The large-scale modes of climate variability, such as
El~Ni\~no-Southern Oscillation (ENSO) and the Pacific decadal
oscillation (PDO), have profound impacts on the precipitation
regime over North America, especially during the
winter [e.g., \citeauthor{Rope:Halp:nort:1986} (\citeyear{Rope:Halp:nort:1986,Rope:Halp:quan:1996}),
\citet{Gers:Barn:enso:1998}].
As an example, El~Ni\~no usually lasts for at least one season and
brings substantially increased
extreme precipitation over a vast region of North America
[\citet{Zhan:Wang:Zwie:Groi:infl:2010,Shan:Yan:Zhan:El:2011}].
Rare events that occur at multiple locations within a very
short time interval can cause more damage, consume more
resources and demand more delicate emergency rescue.
For strategic emergency management and loss mitigation, understanding
and predicting extreme events in a spatial context is needed.
Although univariate extreme value modeling has been well
developed [e.g., \citet{Cole:an:2001}], spatial extreme
modeling has not gained a sharpened focus until recently
[e.g., \citet
{deH:Pere:spat:2006,Bush:deH:Zhou:on:2008,Pado:Riba:Siss:like:2010,Davi:Ghol:geos:2012}].
Two recent reviews are \citet{Davi:Pado:Riba:stat:2012}
and \citet{Bacr:Gaet:revi:2012}; the latter focuses
on spatial max-stable processes, while the former covers
additionally latent variable approaches and copula approaches.

A max-stable process extends the multivariate extreme value
distribution to an infinite dimension; every multidimensional
marginal distribution of it is a multivariate
extreme value distribution [\citet{deH:spec:1984}].
For only a few parametric models, statistical inference
is practically viable:
the Smith model [\citet{Smit:max:1990}],
the Schlather model [\citet{Schl:mode:2002}],
the Brown--Resnick model [\citet{Kabl:Schl:deH:stat:2009}],
the geometric Gaussian model [\citet{Davi:Pado:Riba:stat:2012}]
and the extremal-$t$ model [\citet{Opit:extr:2013}].
\citet{Wads:Tawn:depe:2012} proposed to superimpose two max-stable
processes to obtain a new model, which can produce more realistic
event realizations than, for example, the Smith model by itself.
Inferences for max-stable process models are challenging
because the joint density for multiple sites is only available
for bivariate or trivariate marginal distributions.
In fact, the trivariate marginal density has been derived
only recently for the Smith model [\citet{Gent:Ma:Sang:on:2011}]
and the Brown--Resnick model [\citet{Huse:Davi:comp:2013}].
The pairwise likelihood approach based on the bivariate marginal
distributions of block maxima has been used in applications
[\citet{Pado:Riba:Siss:like:2010,Davi:Ghol:geos:2012}].

For univariate extreme value analysis based on generalized
extreme value (GEV) distributions, daily records which
contain more information than annual maxima can be exploited.
Two well-known threshold approaches are the peaks over threshold (POT)
approach [\citet{Balk:deH:resi:1974,Pick:stat:1975}] and the point
process approach [\citet{Pick:two-:1971,Lead:Root:Lind:Root:extr:1983}].
\citet{Ferr:deH:on:2014} recently showed that, for the probability
weighted moment estimator [\citet{Hosk:Wall:Wood:esti:1985}],
the block maximum method is asymptotically more efficient in
mean squared error than the POT method under certain conditions.
Nonetheless, for shorter records, the threshold methods
may be more efficient than the block maximum method when
the number of exceedances is larger than the number of
blocks on average or when the shape parameter is positive
[e.g., \citet{Katz:Parl:Nave:stat:2002,Tana:taka:a:2002}].
For multivariate extremes, as \citet{Falk:Mich:test:2009} pointed
out, an extension of the univariate threshold approach needs
to solve which distributions describe the exceedances and
how exceedances are defined in a multivariate setting.
These problems are being actively investigated [e.g., \citet
{Root:Tajv:mult:2006,Falk:Guil:peak:2008,Falk:Huss:Reis:laws:2010,Thib:Opit:effi:2013,Ferr:deHa:gene:2014}].
Bivariate threshold-based inferences have been applied to
max-stable process models through the composite likelihood approach.
\citet{Bacr:Gaet:esti:2014} considered two bivariate exceedance
distributions, one from the tail approximation for bivariate distribution
in \citet{Ledf:Tawn:stat:1996} and the other from the bivariate extension
of a generalized Pareto distribution (GPD) in \citet{Root:Tajv:mult:2006}.
No clear winner of the two approaches was found in a simulation
study, and their performance depends on the spatial dependence level.
Similar approaches have been adopted by \citet{Wads:Tawn:effi:2014} for
spatial extremes and \citet{Huse:Davi:spac:2014} in a space--time setting.

Without resorting to a spatial version of threshold-based approaches,
we propose a two-step approach that utilizes daily records in
addition to block maxima from each site for max-stable process models.
The first step estimates the marginal parameters using an
independence likelihood constructed from the univariate
threshold-based point process approach with daily records.
Given the marginal parameter estimates, the second step estimates
the dependence parameters from a pairwise likelihood with block maxima.
The two-step approach has been studied recently for multivariate
models to overcome the computational difficulty in maximum likelihood
estimation [\citet{Zhao:Joe:comp:2005,Joe:asym:2005}].
Our two-step approach is different, however, in that
we use different data in the two steps: the first step uses
daily records, while the second step uses block maxima.
Compared to the bivariate threshold-based approaches, the
marginal parameter estimator from the two-step approach is
robust to misspecification of the spatial dependence.
The more efficient marginal estimator helps improve
the efficiency of the dependence parameter estimator compared
to the composite likelihood estimator based on only block maxima.

The rest of the article is organized as follows.
Our motivating application, annual maximum winter daily
precipitation in California, is presented in Section~\ref{sec:cali}.
The spatial max-stable process model defined by all
univariate marginal distributions and a spatial dependence
structure, and the dependence measure extremal coefficient,
are introduced in Section~\ref{sec:maxstab}.
In Section~\ref{sec:2step} we present details of the
two-step approach, the asymptotic properties of the estimator,
and describe how to estimate the limiting variance.
A simulation study is reported in Section~\ref{sec:simu}.
The proposed method is applied to the precipitation data from
36 sites in California over 55 years in Section~\ref{sec:appl},
providing more compact confidence regions for joint return levels.
Section~\ref{sec:disc} concludes with some discussion.

%s2 #&#
\section{Extreme winter precipitation in California}
\label{sec:cali}

Recent studies suggest that the El~Ni\~no/Southern Oscillation
(ENSO) has significant impact on extreme precipitation in
North America [\citet{Zhan:Wang:Zwie:Groi:infl:2010}].
Southern Oscillation refers to the variation in the sea surface
temperature of the tropical waters in the eastern Pacific Ocean.
The ``warm'' events and the ``cool'' events are referred to
as El~Ni\~no and La~Ni\~na, respectively, and their strength is
measured by the Southern Oscillation Index (SOI), the normalized
sea level pressure difference between Tahiti and Darwin.
With SOI as a covariate in sitewise GEV modeling, El~Ni\~no
was found to be associated with a substantial increase in the
likelihood of extreme precipitation over a vast region of
southern North America [\citet{Zhan:Wang:Zwie:Groi:infl:2010}].
Focusing on the California stations, \citet{Shan:Yan:Zhan:El:2011}
reported similar findings with spatial dependence incorporated
through a Smith model, which enabled inference and predictions
of joint extremal events at multiple sites within the same year.
Nevertheless, two practical issues were not satisfactorily addressed.
First, realizations from the Smith model are of too regular shape.
Second, collaborators who are familiar with threshold-based
univariate extreme value analysis wondered if the full records
of daily precipitation can lead to a more efficient analysis
than that based on block maxima alone.
These issues motivated our two-step approach and a revisit
of the extreme winter precipitation in California.

\begin{figure}

\includegraphics{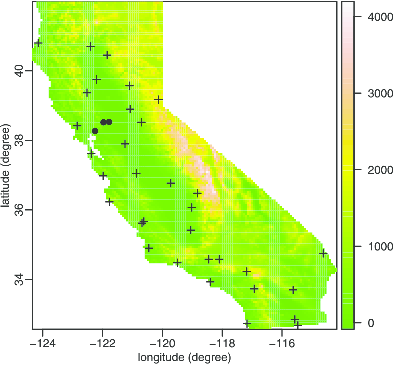}

\caption{Locations of the 36 monitoring stations in California
superimposed with the elevation map (meter).
The three sites in solid circles are Napa, Winters and Davis, near the
Sacramento area.}\label{fig:CAelevmap}
\end{figure}

Daily precipitation records at all monitoring stations in
California were extracted from the second version of the
Global Historical Climatology Network (GHCN), compiled and
quality-controlled at the National Climatic Data Center of
the National Oceanic and Atmospheric Administration (available
at \surl{http://www.ncdc.noaa.gov/oa/climate/ghcn-daily/}).
As precipitation in California occurs predominantly in winter,
we restrict our attention to the winter season, which
is defined as the period from December 1st to March 31st in
the following year [\citet{Zhan:Wang:Zwie:Groi:infl:2010}].
Due to missing data, the block maxima in a given winter
at a given site was considered to be valid only
if no more than 10\% of the daily records were missing
in that winter [\citet{Shan:Yan:Zhan:El:2011}].
For comparison, we used the same time periods and sites as
the balanced data in \citet{Shan:Yan:Zhan:El:2011}, covering
daily winter precipitation from 1948 to 2002 for 36 sites.
The 36 sites in California are shown in Figure~\ref{fig:CAelevmap},
superimposed with the elevation map of the state.
The distance between the two furthest sites is 1188 km.
As in \citet{Shan:Yan:Zhan:El:2011}, possible covariates to be
included in the GEV parameters for each site are
longitude, latitude, elevation and SOI.
The latitude and longitude are in degrees,
and the elevation is in 100 meters.
The SOI for each winter is the average of the four monthly
SOI values of the winter months, ranging from $-3.14$ to
$1.88$ with a sample average $-0.15$ for the data period.

%f1 #&#

%s3 #&#
\section{Spatial extreme model with max-stable process}
\label{sec:maxstab}

A max-stable process model for spatial extremes consists of two
parts: marginal distributions and spatial dependence structure.
The marginal distribution at each site is a GEV distribution,
which may incorporate temporal nonstationarity through
temporally varying covariates such as the SOI.
In particular, let $M(s,t)$ be the maximum at site
$s$ in block $t$ in a spatial domain $D \subset\mathbb{R}^2$.
The distribution of $M(s,t)$ is
%
%e3.1 #&#
\begin{equation}
\label{eq:marg} M(s,t) \sim\operatorname{GEV} \bigl(\mu(s,t), \sigma(s,t), \xi(s,t)
\bigr),
\end{equation}
where $\mu(s,t)$, $\sigma(s,t)$ and $\xi(s,t)$ are the location,
scale and shape parameters, respectively, of the GEV distribution.
Covariate information is incorporated into the parameters through
% \begin{eqnarray*}
$\mu(s,t) = X^{\top}_{\mu}(s,t)\beta_{\mu}$,
$\sigma(s,t) = X^{\top}_{\sigma}(s,t) \beta_{\sigma}$,
and $\xi(s,t) = X^{\top}_{\xi}(s,t) \beta_{\xi}$,
% \end{eqnarray*}
where $X_{\mu}(s,t)$, $X_{\sigma}(s,t)$ and $X_{\xi}(s, t)$ are the
covariate vectors for $\mu$, $\sigma$ and $\xi$, respectively,
$\top$ denotes transpose, and
$\beta= (\beta^{\top}_{\mu}, \beta^{\top}_{\sigma}, \beta^{\top}_{\xi
})^{\top}$
is the vector containing all marginal parameters.

The spatial dependence structure ensures that every finite-dimensional
marginal distribution is a multivariate GEV distribution.
The multivariate extreme value property essentially requires
that every finite-dimensional marginal copula must be
an extreme value copula [\citet{Gude:sege:extr:2010}].
Without loss of generality, let
$Z(s,t) = F^{-1}[G_{s,t}\{( M(s,t) \}]$,
where $F$ is the distribution function of a unit
Fr\'echet variable with inverse function $F^{-1}$, and
$G_{s,t}(\cdot; \beta)$ is the distribution function of
$\operatorname{GEV} (\mu(s,t), \sigma(s,t), \xi(s,t) )$
with parameter vector $\beta$.
Consider any $p$ sites $x_i \in D$, $i = 1, \ldots, p$.
The copula of $\{ M(x_1,t), \ldots, M(x_p,t) \}$ is the
same as the copula of $\{ Z(x_1,t), \ldots, Z(x_p,t) \}$.
This copula is determined by a max-stable process (MSP) model
with dependence parameter $\theta$ for process $Z(s,t)$
for any $t$:
%
%e3.2 #&#
\begin{equation}
\label{eq:dep} Z(s, t) \sim\operatorname{MSP}(\theta).
\end{equation}
The MSP has a marginal unit Fr\'echet distribution at each $s$
and marginal extreme value copulas for any multidimensional
marginal distribution.

%% max-stable processes models
The parametric form of an MSP is determined from its spectral
representation [\citet{deH:spec:1984,Schl:mode:2002}].
Let $\{U_j\}_{j \geq1}$ be a Poisson process on
$\mathbb{R}_+$ with intensity $\dif u / u^2$.
Let $W_{j} (x)$, $x \in D$, $j \geq1$, be independent
copies of a nonnegative stationary process $W(x)$ with
$E\{W(x)\} = 1$ for all $x \in D$.
Then,
\[
\label{eq:spec} Z(x) = \sup_{j \geq1} U_j
W_j(x),  \qquad  x \in D,
\]
is a stationary MSP with unit Fr\'echet margins.
Three practically viable MSP models are obtained by
different choices of $W(x)$ with parameter vector $\theta$
[e.g., \citet{Davi:Pado:Riba:stat:2012}].
The Smith model takes $W_j(x) = g(x - V_j)$, where
$g$ is the density of a zero mean bivariate normal random vector
with variance matrix $\Sigma$, and $V_1, V_2, \ldots$ are
the points of a homogeneous Poisson process of unit rate in $D$.
Isotropy is obtained when $\Sigma= \tau I_2$, where $I_2$ is
the two-dimensional identity matrix and $\tau> 0$ is a scalar.
The Schlather model takes
$W_j(x) = \max\{0, \sqrt{2 \pi} \varepsilon_j(x)\}$, where
$\varepsilon_1,\varepsilon_2,\ldots$ are independent copies of
a stationary Gaussian process $\{\varepsilon(x) \dvtx x \in D \}$
with unit variance and correlation function $\rho$.
A geometric Gaussian model takes
$W_j(x) = \exp\{\delta\varepsilon_j(x) - \delta^2/2 \}$,
where $\delta> 0$ and $\varepsilon_j(x)$'s are independent copies
of a stationary Gaussian process with unit variance and
correlation function $\rho$.
Geometric anisotropy can be obtained for the Schlather model and
the geometric Gaussian model through using a geometric anisotropic
correlation function $\rho$.
The spectral representation of another model, the extremal-$t$
process, was only obtained recently by \citet{Opit:extr:2013}.
The \mbox{extremal-$t$} process is the extreme value limit of $t$
processes, which are scale mixtures of Gaussian processes.
It is characterized by a degree of freedom $\nu$ and a dispersion
function $\rho$ (the correlation function of the Gaussian process).
The extremal-$t$ process covers the Schlather model when $\nu= 1$
and the Brown--Resnick model when $\nu\to\infty$.

%% extremal coefficient
A useful measure for extremal dependence is the extremal coefficient.
For an MSP $Z(s)$ with unit Fr\'echet margins, the extremal coefficient
at $p$ sites $x_1, \ldots, x_p$ is the number $\zeta$ such that
\[
\label{eq:extremecoeff} \Pr\bigl\{Z(x_{1})\leq z, \ldots, Z(x_{p})
\leq z \bigr\} = \exp(-\zeta/ z), \qquad  z > 0.
\]
The range of $\zeta$ is $1 \leq\zeta\leq p$, with 1 and $p$
corresponding to full dependence and independence, respectively.
The pairwise extremal coefficient as a function of the pairwise
distance can be used in exploratory analysis and model checking.
For two sites $x_1$ and $x_2$ with $h = x_2 - x_1$,
the pairwise extremal coefficient $\zeta(h)$ is
$2 \Phi\{ \sqrt{h^{\top} \Sigma^{-1} h}/2 \}$,
$1 + \sqrt{[1 - \rho(h)] / {2}}$,
$2 \Phi\{ \sqrt{\delta^2 [1 - \rho(h)]/2} \}$,
and $2 T_{\nu+1} \{\sqrt{[1-\rho(h)][\nu+1]/[1+\rho(h)]}\}$
for the Smith, Schlather, geometric Gaussian and extremal-$t$ model,
respectively, where $\Phi$ is the distribution function of
a standard normal variate and $T_{\nu}$ is the distribution
function of a Student-$t$ variate with $\nu$ degrees of freedom.
Unlike the other three models which offer the full range of
dependence level from complete independence to complete dependence,
the Schlather model has
$\zeta(h) \le1 + \sqrt{1/2} \approx1.707$, not allowing
full independence of two sites regardless of their distance.

%% pairwise likelihood
Given observed block maxima from $S$ sites over $n$ years,
estimation of model parameters
$\eta= (\beta^{\top}, \theta^{\top})^{\top}$
is challenging because the full joint distribution of $S$ sites
is unavailable for $S \ge3$ in general.
Inference about max-stable process models has mostly been based on the
composite likelihood approach
[e.g., \citet{Pado:Riba:Siss:like:2010,Davi:Ghol:geos:2012}].
In particular, a~pairwise likelihood is constructed from
the bivariate marginal distributions of all pairs.
The three aforementioned MSP models are viable because their bivariate
marginal distributions have closed forms and the corresponding
density can be derived and used to construct pairwise likelihoods.
The pairwise likelihood approach is potentially wasteful of data
because it only uses the block maxima.

%s4 #&#
\section{The two-step approach}
\label{sec:2step}

Suppose that we observe the full record of each block with block
size $m$ at $S$ sites over $n$ blocks (e.g., years or seasons).
For ease of notation, $m$ is assumed to be the same but our approach
can also handle the case where $m$ varies from year to year.
Let $Y_{s,t,k}$ be the $k$th observation within block $t$ at
site $s$, $k=1, \ldots, m$, and let
$\mathbf{Y_{s,t}} = \{Y_{s,t,1}, \ldots, Y_{s,t,m}\}$.
Let $M_{s,t} = \max_k Y_{s,t,k}$ be the block maximum.
Our first step estimates the marginal parameters $\beta$
based on daily records
$\mathbf{Y} = \{\mathbf{Y}_{s,t}\dvtx s = 1, \ldots, S;  t=1, \ldots,
n\}$.
Our second step estimates the dependence parameters $\theta$
based on block maxima
$\mathbf{M} = \{M_{s,t}\dvtx s = 1, \ldots, S;  t=1, \ldots, n\}$.

\emph{Step} 1.
The first step is based on an independence likelihood constructed from
the point process approach for univariate extreme value analysis.
This step utilizes the daily record in each block
but ignores the spatial dependence across sites.
Let $u_{s,t}$ be the threshold chosen for site $s$ and block
$t$, $s = 1, \ldots, S$, $t = 1, \ldots, n$.
This choice accommodates nonstationarity across the blocks.
The independence loglikelihood has the form
%
%e4.1 #&#
\begin{equation}
\label{eq:l1} l_1(\beta; \mathbf{Y}) = \sum
_{t=1}^n \sum_{s=1}^S
\ell_{1t,s}(\beta; \mathbf{Y}_{s,t}),
\end{equation}
where
\begin{eqnarray*}
&&\ell_{1t,s}(\beta; \mathbf{Y}_{s,t})\\
&&\qquad = - \biggl[1 +
\xi_{s,t} \biggl(\frac{u_{s,t} - \mu_{s,t}}{\sigma_{s,t}} \biggr) \biggr]^{-1/\xi
_{s,t}}
\\
&&\qquad\quad{}+ \sum_{k: Y_{s,t,k} > u_{s,t}} \biggl[ -\log\sigma_{s,t} -
\biggl(\frac{1}{\xi_{s,t}} + 1 \biggr) \log \biggl\{1 + \xi_{s,t} \biggl(
\frac
{Y_{s,t,k}-\mu_{s,t}}{\sigma_{s,t}} \biggr) \biggr\} \biggr].
\end{eqnarray*}
The contribution to the independence loglikelihood from site $s$,
$\sum_{t=1}^n \ell_{1t,s}$, is simply the loglikelihood of the point process
approach in a univariate extreme value analysis [\citet{Smit:extr:1989}].
Since we assume independence from block to block, the contribution
from block $t$ is $\ell_{1t} = \sum_{s=1}^S \ell_{1t,s}$.
The maximizer of~\eqref{eq:l1}, $\hat\beta_n$, is the estimator of
$\beta$.

The independence loglikelihood~\eqref{eq:l1} also allows temporal
dependence within the same block, in which case the temporal
dependence is ignored similar to the spatial dependence.
Recent studies show that this approach not only uses all
threshold excesses for more efficient estimation, but also
avoids significant biases that may come with declustering
[\citeauthor{Fawc:Wals:impr:2007}
(\citeyear{Fawc:Wals:impr:2007,Fawc:Wals:esti:2012})].
The variance of $\hat\beta_n$ needs to be estimated with sandwich
estimators to adjust for the dependence [\citet{Smit:Regi:1991}].

\emph{Step} 2.
Given $\hat\beta_n$, the second step uses block maxima to estimate
the dependence parameters $\theta$ based on a pairwise likelihood.
Let $f_{ijt}(\cdot; \theta, \beta)$ be the bivariate marginal density
of the $(M_{i,t}, M_{j,t})$ from the max-stable process model specified
by~\eqref{eq:marg} and~\eqref{eq:dep} with site $i$ and $j$ in block $t$.
Define pairwise loglikelihood
%
%e4.2 #&#
\begin{equation}
\label{eq:l2} l_2(\theta; \hat\beta_n, \mathbf{M}) =
\sum_{t=1}^{n}\ell_{2t}(\theta;
\hat\beta_n, M_{s,t}\dvtx s = 1, \ldots, S),
\end{equation}
where the contribution from block $t$ is
\[
\ell_{2t}(\theta; \beta, M_{s,t}\dvtx s = 1, \ldots, S) =
\sum_{i=1}^{S-1} \sum
_{j=i+ 1}^{S} \log f_{ijt}
\bigl((M_{i,t}, M_{j,t}); \theta, \beta \bigr).
\]
Our estimator for $\theta$, $\hat\theta_n$,
is the maximizer of~\eqref{eq:l2}.

The asymptotic properties of the two-step estimator
$\hat\eta_n^{\top} = (\hat\beta_n^{\top}, \hat\theta_n^{\top})$ can be
derived with the theory of estimating functions [\citet{Goda:esti:1991}].
Let $\psi_{1t}(\beta) = \partial\ell_{1t} /\partial\beta$.
Let $\psi_{2t}(\beta, \theta) = \partial\ell_{2t} / \partial\theta$.
Then $\hat\eta_n$ is the solution to the estimating equations
$\sum_{t=1}^n \psi_t (\eta) = 0$,
where
$ \psi_t^{\top}(\eta) =
 (\psi_{1t}^{\top}(\beta), \psi_{2t}^{\top}(\beta, \theta) )$.
Under mild regularity conditions, as $n \to\infty$,
$\hat\eta_n$ is consistent for the true parameter vector $\eta_0$, and
$\sqrt{n}(\hat\eta_n - \eta_0) \to N(0, \Omega)$,
where $\Omega= A^{-1} B (A^{-1})^{\top}$ is the inverse of the
Godambe information matrix, with
$ A = \lim_{n \to\infty} n^{-1} \sum_{t=1}^n \partial\psi_t(\eta) /
\partial\eta^{\top}$
and
$ B = \lim_{n \to\infty} n^{-1} \sum_{t=1}^n \psi_t(\eta) \psi_t^{\top
}(\eta)$.
With independent replicates at the block level, $\Omega$~can be
easily estimated with the sample versions of $A$ and $B$ as outlined
in the supplementary material [\citet{Shan:Yan:Zhan:supp:2014}].
An alternative, computing-intensive method is a bootstrap
applied to the blocks (years) with spatial structure preserved.
We assess the validity of the sandwich estimator in our simulation
study but use the bootstrap estimator in the real data analysis.

Computationally, the optimization in both steps can be challenging,
especially when the dimension of the parameter vector is large.
Optimizing with respect to all parameters simultaneously often
gives poor results at local maxima [\citet{Blan:Davi:spat:2011}].
We adapt the profile method suggested for pairwise likelihood
maximization by \citet{Blan:Davi:spat:2011} and apply it to both steps.
The profile method maximizes with respect to one parameter at a time
while holding all other parameters at their current values, and the
process goes through all parameters iteratively until convergence.
To be safe, we optimize with respect to all parameters simultaneously
one more time after the convergence of the profile method.

Model selection for the two-step approach can be done separately
for the marginal GEV models and the MSP model in two steps with
the composite likelihood information criterion (CLIC)
[\citet{Vari:Vido:note:2005,Vari:on:2008}], which is an adaptation of
the Takeuchi information criterion (TIC) [\citet{Take:dist:1976}].
Models with lower $\operatorname{CLIC}$ are preferred.
In step~1, the CLIC selects the best marginal model
without specifying the spatial dependence structure.
In step~2, the CLIC is a conditional version given the marginal
model selected from step~1 and the marginal parameter estimates.

%s5 #&#
\section{Simulation study}
\label{sec:simu}

To investigate the performance of the two-step approach using
daily records in comparison to the pairwise likelihood approach
using block maxima only, a simulation study was conducted.
The study region was confined to $[-20, 20]^2$.
The marginal distribution of the block maxima at each site $s$ is a GEV
distribution with location $\mu_s$, scale $\sigma_s$ and shape $\xi_s$.
Let $X_1(s)$ and $X_2(s)$ denote the latitude and longitude of site $s$.
The GEV parameters were
\[
\label{eq:simulmargmodel} %
\cases{ \mu_s = \beta_{\mu,0}
+ \beta_{\mu,1} X_1(s) + \beta_{\mu,2}
X_2(s),
\vspace*{2pt}\cr
\sigma_s = \beta_{\sigma,0} + \beta_{\sigma,1}
X_1(s) + \beta_{\sigma
,2} X_2(s),
\vspace*{2pt}\cr
\xi_s = \beta_{\xi,0},}
\]
where $\beta_{\mu,0} = 15$, $\beta_{\mu,1} = -0.2$, $\beta_{\mu,2} = 0.25$,
$\beta_{\sigma,0} = 4$, $\beta_{\sigma,1} = -0.04$, $\beta_{\sigma,2}
= 0.08$, and $\beta_{\xi,0} = 0.2$.
The factors of our simulation study are as follows:
the max-stable model, the spatial dependence level,
the number of sites $S$ and the sample size $n$.
Three one-parameter isotropic max-stable processes were considered:
the Smith model, the Schlather model and the geometric Gaussian model.
The Smith model has a single parameter $\theta= \tau$.
The Schlather model has an exponential correlation function with
range parameter $\theta= \alpha$:
$\rho(h) = \exp( - \| h \| / \alpha)$.
The geometric Gaussian model also has an exponential correlation
function with range parameter $\theta= \alpha$ and the
parameter $\delta^2 = 8$ is assumed known.
The choice of the value 8 is a compromise between two facts:
(1) the random number generation from this model in R package
\texttt{SpatialExtremes} [\citet{Rpkg:SpatialExtremes}] works well
only for $\delta^2 < 10$; and (2) the pairwise extremal coefficient
from $\delta^2 = 8$ has an upper bound 1.96, close to independence.
The Brown--Resnick model, which covers the geometric Gaussian
model as a special case and offers full range of dependence level,
was not considered here because of lack of fast simulation tools.
Three dependence levels were considered: weak, moderate and
strong, abbreviated as W, M and S, respectively.
The parameter $\tau$ for the Smith model was chosen to
be 20, 200, 2000 for weak, moderate and strong dependence,
respectively, as in \citet{Pado:Riba:Siss:like:2010}.
The range parameters for the other three models were chosen such that
their pairwise extremal coefficient as a function of distance matches
as closely as possible with that from the corresponding Smith model.
From nonlinear least squares fits with distance in the range of
$[0, 40]$, the parameter values of $\alpha$ for the Schlather model
were found to be 5.2, 24.3 and 242.9 for dependence level W, M and S,
respectively, and the corresponding $\alpha$ values for the geometric
Gaussian model were found to be 25.2, 135.2 and 1252.0, respectively.
We considered two levels for the number of sites $S \in\{25, 50\}$
and three levels for sample size $n \in\{20, 50, 100\}$.
The performance of the sandwich variance estimator for $n = 20$ was not
expected to be good, but we kept $n = 20$ in efficiency comparisons.

For each scenario, 1000 data sets of daily records were generated.
The $S$ sites were regenerated for each data set from
a uniform distribution over the study region $[-20, 20]^2$.
To mimic the California data analysis, we set block size $m = 122$.
The $m$ daily observations at $S$ sites within each season were
generated as realizations from the target MSP model divided by $m$.
The max-stability ensures that the site-wise maxima of the $m$
observations at the $S$ sites is a realization from the MSP model.
For each data set, we used the profile method for both approaches
with the same starting values---the pairwise likelihood estimate
from R package \texttt{SpatialExtremes} [\citet{Rpkg:SpatialExtremes}].
The threshold $u_{s,t}$ in the two-step approach was chosen to be the
95th sample percentile at site $s$ in block $t$.

We first assess the estimator from the two-step approach.
The results for $n \in\{50, 100\}$ are summarized in tables
in the supplementary material [\citet{Shan:Yan:Zhan:supp:2014}].
Consider, for example, the geometric Gaussian model.
The biases are very small relative to the truth for all parameters.
The empirical standard error of the estimates is higher for
stronger dependence or smaller sample size, but it is much
less sensitive to the number of sites $S$, which is consistent
with the observation in \citet{Pado:Riba:Siss:like:2010}.
The average standard errors are generally in close agreement
with the empirical standard errors, suggesting good performance
of the sandwich variance estimator for sample size as small as 50.
Consequently, the empirical coverage percentage of the 95\%
confidence intervals for most parameters are close to the nominal level.
Under-coverage occurred for $\alpha$ and $\beta_{\xi,0}$ when the
dependence is weak; the lowest case was 84\% for $S = 50$ and $n = 100$.
The coverage for $\log\alpha$ is uniformly better than for $\alpha$.
The under-coverage is unfortunate because sandwich variance estimators
tend to underestimate the variance for small to moderate sample sizes.
Bias correction [\citet{Manc:DeRo:cova:2001,Kaue:Carr:note:2001}]
might lead to better coverage rate of the confidence intervals
in this context, but an investigation is beyond our scope here.
The results for the Smith model and the Schlather model were
similar or better---no empirical coverage was below 90\%.

%t1 #&#
\begin{sidewaystable}
\tabcolsep=0pt
\tablewidth=\textwidth
\caption{Relative efficiency (\%) in mean squared error of model
parameter estimates for the pairwise likelihood approach
relative to the two-step approach for Smith,
Schlather and geometric Gaussian models}
\label{tab:relaeffimodel1}
\begin{tabular*}{\textwidth}{@{\extracolsep{\fill}}ld{3.0}cccccccccd{3.0}ccccccccccccccc@{}}
\hline
& & & \multicolumn{8}{c}{\textbf{Smith}} &
\multicolumn{8}{c}{\textbf{Schlather}} & \multicolumn{8}{c@{}}{\textbf{Geometric Gaussian}} \\[-6pt]
& & & \multicolumn{8}{c}{\hrulefill} &
\multicolumn{8}{c}{\hrulefill} & \multicolumn{8}{c@{}}{\hrulefill} \\
\textbf{Dep} & \multicolumn{1}{c}{$\bolds{n}$} & $\bolds{S}$ &
$\bolds{\tau}$ & $\bolds{\beta_{\mu,0}}$ & $\bolds{\beta_{\mu,1}}$ & $\bolds{\beta_{\mu,2}}$ & $\bolds{\beta
_{\sigma,0}}$ & $\bolds{\beta_{\sigma,1}}$ & $\bolds{\beta_{\sigma,2}}$ & $\bolds{\beta_{\xi
,0}}$ &
\multicolumn{1}{c}{$\bolds{\alpha}$} & $\bolds{\beta_{\mu,0}}$ & $\bolds{\beta_{\mu,1}}$ & $\bolds{\beta_{\mu,2}}$ & $\bolds{\beta
_{\sigma,0}}$ & $\bolds{\beta_{\sigma,1}}$ & $\bolds{\beta_{\sigma,2}}$ & $\bolds{\beta_{\xi
,0}}$ &
$\bolds{\alpha}$ & $\bolds{\beta_{\mu,0}}$ & $\bolds{\beta_{\mu,1}}$ & $\bolds{\beta_{\mu,2}}$ & $\bolds{\beta
_{\sigma,0}}$ & $\bolds{\beta_{\sigma,1}}$ & $\bolds{\beta_{\sigma,2}}$ & $\bolds{\beta_{\xi
,0}}$ \\
\hline
W & 20 & 25 & 81 & 67 & 54 & 54 & 72 & 25 & 24 & 30 & 107 & 66 & 64 &
62 & 87 & 36 & 48 & 32 & 87 & 75 & 56 & 57 & 86 & 25 & 30 & 37 \\
& & 50 & 81 & 70 & 54 & 59 & 80 & 24 & 30 & 34 & 106 & 64 & 65 & 62 &
92 & 41 & 57 & 34 & 83 & 87 & 57 & 60 & 92 & 29 & 34 & 45 \\
& 50 & 25 & 80 & 70 & 55 & 57 & 77 & 23 & 28 & 33 & 98 & 71 & 60 & 61
& 87 & 34 & 49 & 37 & 85 & 82 & 58 & 60 & 89 & 24 & 32 & 43 \\
& & 50 & 78 & 76 & 55 & 62 & 89 & 23 & 32 & 35 & 100 & 66 & 66 & 65 &
87 & 40 & 54 & 40 & 85 & 85 & 55 & 58 & 95 & 25 & 35 & 46 \\
& 100 & 25 & 81 & 73 & 56 & 57 & 80 & 24 & 28 & 34 & 97 & 67 & 61 & 62
& 91 & 34 & 49 & 37 & 85 & 84 & 52 & 60 & 94 & 23 & 34 & 45 \\
& & 50 & 74 & 75 & 57 & 60 & 85 & 25 & 30 & 37 & 101 & 77 & 64 & 72 &
97 & 40 & 61 & 36 & 87 & 89 & 55 & 57 & 96 & 28 & 34 & 55 \\
M & 20 & 25 & 84 & 62 & 51 & 55 & 77 & 29 & 37 & 40 & 93 & 58 & 66 &
64 & 81 & 44 & 54 & 36 & 85 & 63 & 49 & 49 & 86 & 28 & 39 & 54 \\
& & 50 & 90 & 63 & 51 & 56 & 78 & 30 & 35 & 40 & 96 & 63 & 66 & 66 &
86 & 42 & 55 & 38 & 77 & 67 & 49 & 55 & 80 & 28 & 43 & 53 \\
& 50 & 25 & 86 & 69 & 49 & 57 & 78 & 27 & 35 & 45 & 94 & 59 & 65 & 61
& 80 & 42 & 48 & 44 & 80 & 66 & 52 & 51 & 80 & 29 & 38 & 52 \\
& & 50 & 86 & 64 & 53 & 52 & 80 & 33 & 38 & 45 & 92 & 65 & 66 & 64 &
84 & 42 & 53 & 41 & 82 & 72 & 52 & 54 & 89 & 29 & 42 & 58 \\
& 100 & 25 & 87 & 69 & 51 & 57 & 82 & 29 & 39 & 49 & 95 & 62 & 75 & 64
& 81 & 42 & 49 & 40 & 79 & 70 & 50 & 55 & 84 & 31 & 43 & 56 \\
& & 50 & 84 & 68 & 50 & 51 & 79 & 30 & 38 & 45 & 93 & 66 & 63 & 69 &
85 & 39 & 58 & 40 & 83 & 70 & 48 & 56 & 89 & 29 & 43 & 60 \\
S & 20 & 25 & 78 & 54 & 58 & 54 & 67 & 47 & 48 & 54 & 79 & 55 & 86 &
67 & 69 & 65 & 67 & 45 & 76 & 55 & 53 & 52 & 66 & 44 & 54 & 78 \\
& & 50 & 69 & 49 & 56 & 50 & 62 & 39 & 49 & 56 & 85 & 55 & 85 & 67 &
67 & 67 & 62 & 52 & 84 & 57 & 57 & 51 & 69 & 48 & 54 & 85 \\
& 50 & 25 & 76 & 59 & 55 & 61 & 69 & 48 & 53 & 56 & 82 & 55 & 81 & 63
& 62 & 60 & 62 & 48 & 75 & 57 & 53 & 57 & 64 & 43 & 54 & 85 \\
& & 50 & 84 & 52 & 48 & 52 & 63 & 38 & 50 & 61 & 88 & 56 & 77 & 66 &
70 & 66 & 72 & 45 & 78 & 61 & 54 & 60 & 67 & 45 & 53 & 85 \\
& 100 & 25 & 86 & 57 & 56 & 53 & 66 & 40 & 47 & 70 & 85 & 57 & 78 & 67
& 63 & 57 & 62 & 44 & 76 & 57 & 51 & 52 & 66 & 42 & 51 & 84 \\
& & 50 & 82 & 63 & 58 & 58 & 68 & 43 & 54 & 68 & 87 & 63 & 85 & 74 &
66 & 59 & 63 & 47 & 74 & 62 & 55 & 58 & 70 & 45 & 54 & 82 \\
\hline
\end{tabular*}
%
%% \end{table}
\end{sidewaystable}

We now compare the efficiency of the pairwise likelihood approach
using block maxima only ($M_1$) with the two-step approach ($M_2$).
Table~\ref{tab:relaeffimodel1} reports the relative efficiency
in mean squared error for the estimators from the two approaches
for each parameter, with the $M_2$ estimator as the reference.
Method $M_2$ has smaller MSE for all marginal parameters;
the relative efficiency of $M_1$ ranges from 23\% to 95\%.
For example, for the shape parameter $\beta_{\xi,0}$ in the
geometric Gaussian model, the relative efficiency of $M_1$
was 45\% for $S = 25$ and $n = 100$, which is the case where the
coverage of the confidence interval was low in the two-step approach.
This is of great interest since the shape parameter $\xi$
governs the tail behavior of the GEV distribution and
plays an important role in predicting return levels.
The difference between the two methods decreases as
the dependence level increases from weak to strong.
For the dependence parameters, the relative efficiency of $M_1$
ranges from 69\% to 103\%, with the highest relative efficiency
occurring in the weak dependence case under the extremal t process.
The efficiency gain in $M_2$ here is explained by the fact that the
marginal parameters are estimated more precisely in the first step.

How does the efficiency gain in $M_2$ affect risk analysis
such as estimation of joint and individual return levels?
Let $y_{50}$ be the joint 50-year return level for two sites $s_1$
and $s_2$, such that $\Pr (Y(s_1) > y_{50}, Y(s_2) > y_{50} ) = 1/50$.
Given the bivariate marginal distribution, $y_{50}$
can be found numerically for any given parameter vector.
We considered three sites in the study region,
$s_1 = (10, 10)$, $s_2 = (10, 11)$, and $s_3 = (10, 0)$.
The joint 50-year return level was estimated for two pairs,
$(s_1, s_2)$ and $(s_1, s_3)$, which represent pairs that
are close and distant, respectively.
The relative efficiency of the two methods in estimating the
individual 50-year return level at the three sites and the joint
50-year return level at the two sites is summarized in a table
in the supplementary material [\citet{Shan:Yan:Zhan:supp:2014}].
The relative efficiency of $M_1$ with $M_2$ as the reference is
poor, ranging from 58\% to 91\% when the dependence is weak.
As the dependence level gets stronger, $M_1$ becomes almost
as competitive as $M_2$, which is consistent with the relative
efficiency for the shape parameter estimator.
The sample size $n$ and the number of sites $S$ seems to
have little effect on the relative efficiency for all three models.

Up to this point, both the GEV margin model and the max-stable
dependence model have been correctly specified in the fitting.
The only possible misspecification for the two-step approach
is the distribution of the exceedances over the threshold,
which depends on the block size $m$ and the threshold $u$.
In practice, however, neither the marginal model nor the
dependence model will be correct for any finite $m$ or $u$,
which may introduce bias in estimation.
To understand the limitation of the two-step method, we
generated data using $t$ processes, which are in the max-domain
of attraction of the extremal-$t$ process [\citet{Opit:extr:2013}].
Details about the data generation, the choice of degree of
freedom $\nu$ and the results for $\nu\in\{1, 2\}$ are in the
supplementary material [\citet{Shan:Yan:Zhan:supp:2014}].
The two-step method was more efficient than the pairwise
likelihood method in all parameter estimation except in
a very few parameters, including $\beta_{\xi,0}$ when $\nu= 2$.
A~close examination revealed that the MSE for the two-step
estimator was dominated by its bias in these cases.
The pairwise likelihood approach requires the convergence
of the marginal block maximum to a $\nu$-Fr\'echet distribution, with
$\nu$ being the degrees of freedom of the $t$ process, and the
convergence of the dependence structure to extremal-$t$ copula.
The two-step approach requires additionally that the distribution of
those observations exceeding the threshold converges to a generalized
Pareto distribution with appropriately transformed parameters.
For $\nu= 1$, the limiting distribution provided good approximation
in all aspects, but for $\nu= 2$, the convergence of the marginal block
maxima and exceedances needed $m$ to be much greater than 122.
Consequently, the two-step method was more efficient than the pairwise
likelihood method for $\nu= 1$, but lost its edge for some parameters
for $\nu= 2$.

% In each season, $m$ independent realizations were generated
% from a centered $t$ process with a dispersion matrix $\Sigma$
% filled by an exponential correlation function.
% These realizations were appropriately scaled so that the transformed
% site-wise maxima converges to the desired marginal GEV distributions
%and
% the dependence structure converges to that of the extremal-$t$
%process.
% We experimented with various $m$ for several choices of the degrees of
% freedom $\nu$ of the $t$ process, and tested the goodness-of-fit of
% the marginal GEV distribution for the block maxima, GPD distribution
% for the exceedances, and the extremal-$t$ copula for some pairs.

% Simulation was carried out with $\nu\in\{1, 2\}$
% The results indicated that for a very big $m$ is needed to pass the
% Kolmogorov--Smirnov test for the marginal maxima for larger $\nu$.
% So we considered $\nu\in\{1, 2\}$ only and the results are
% summarized in Supplementary Materials \citep{Shan:Yan:Zhan:supp:2014}.
% The two-step method is more efficient than the pairwise
% likelihood method in estimating all parameters when $\nu= 1$.
% For $\nu= 2$, the two-step estimates are still more efficient
% in most parameter estimates, but less efficient than the pairwise
% likelihood estimates in $\xi$.

%s6 #&#
\section{Data analysis}
\label{sec:appl}

%s6.1 #&#
\subsection{First step---marginal GEV models}

Recall that our main interest is to make inferences about the
effect of ENSO on extreme precipitation in California.
Let $X_1(t)$ be the SOI in year $t$.
We considered site specific GEV models with SOI in the
location parameter:
%
%e6.1 #&#
\begin{eqnarray}
\label{eq:CAmarg} \mu(s,t)& = & \beta_{\mu,s,0} + \beta_{\mu,s,1}
X_1(t),
\nonumber
\\
\sigma(s,t) &= & \beta_{\sigma,s,0},
\\
\xi(s,t) &= & \beta_{\xi,s,0}.
\nonumber
\end{eqnarray}
This model has $4S$ parameters, but
it does not assume any smooth surface of the GEV parameters
in covariates such as latitude, longitude and elevation,
which may be unrealistic given the complex terrain of California.
In fact, in our earlier exploratory analysis, including all the
covariates in smooth GEV parameter surfaces led to undesired
results: the effects of the SOI made little physical sense and
the GEV models did not pass goodness-of-fit tests at many sites.

Model~\eqref{eq:CAmarg} was fitted with threshold $u(s,t)$ chosen
to be the 98th sample percentile of the daily records in block
$t$ at site $s$ in the first step of our two-step approach.
The standard errors of the parameter estimates were obtained
by the bootstrap method with 1000 bootstrap samples.
To check the adequacy of the marginal GEV models, a parametric
bootstrap based goodness-of-fit test procedure was performed
for the annual winter maximum daily precipitation at each site.
Out of 36 sites, the $p$-values of the Kolmogorov--Smirnov
test statistics at 35 sites were insignificant at the 1\% level.
The choice of 1\% level was ad hoc and informal, with the
consideration of multiple tests and possible adjustment to
control false discovery rate [e.g., \citet{Benj:Yeku:cont:2001}].
The only site that did not pass the goodness-of-fit test
was removed from the analysis in the sequel.

%f2 #&#
\begin{figure}

\includegraphics{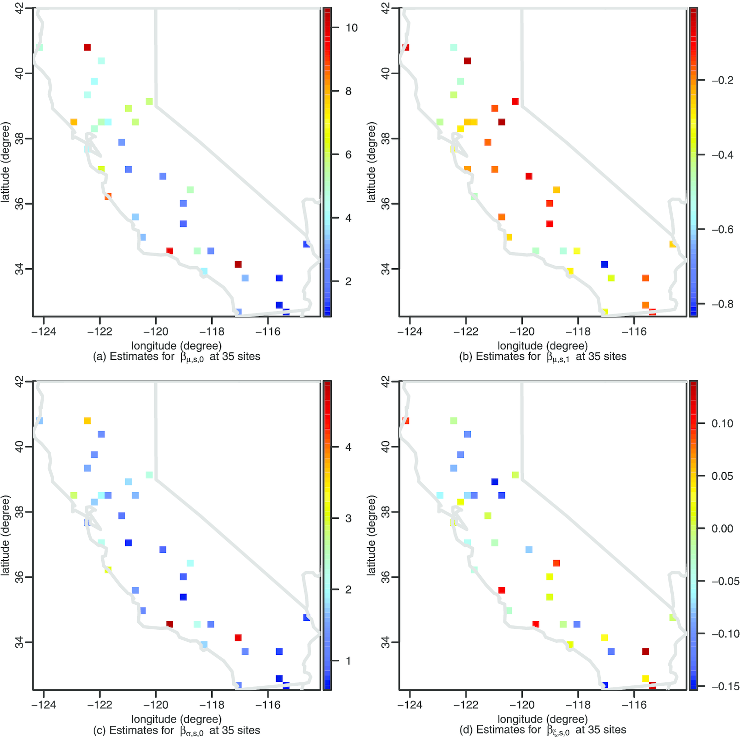}

\caption{Marginal parameter estimates at 35 sites.
\textup{(a)} $\beta_{\mu, s, 0}$;
\textup{(b)} $\beta_{\mu, s, 1}$;
\textup{(c)} $\beta_{\sigma, s, 0}$;
\textup{(d)} $\beta_{\xi, s, 0}$.}
\label{fig:Margcoeff.est.2step}
\end{figure}

The pattern of the marginal parameter estimates at 35 sites
is presented in Figure~\ref{fig:Margcoeff.est.2step}.
It confirms that there is no obvious smooth surface of
these parameters to be characterized by simple functions
of covariates such as latitude, longitude and elevation.
Our interest is the coefficients of SOI in the location parameter.
Their estimates were negative at all sites, and 22 out of
the 35 estimates were significantly negative at the 5\% level.
We also investigated the map of the standardized coefficient
estimates, estimates divided by their standard errors,
in the supplementary material [\citet{Shan:Yan:Zhan:supp:2014}].
The standardized coefficient estimates are the $z$-scores under the
null hypotheses that the corresponding coefficients are zero.
Again, no obvious smooth spatially varying pattern was present.

%s6.2 #&#
\subsection{Second step---spatial dependence model}

Using the fitted marginal GEV models from the first step, we
transformed the block maxima to the unit Fr\'echet scale.
An exploratory analysis with the pairwise extremal coefficients
of the transformed data using the Cap\'era\`a--Foug\`eres--Genest
(CFG) estimator [\citet{Cape:Foug:Gene:nonp:1997,Gene:Sege:rank:2009}]
suggested possible anisotropy and elevation effect in the dependence.
We considered both the Schlather model and the geometric Gaussian
model with a climate space transformation to allow anisotropy and
elevation effects [\citet{Cool:Nych:Nave:baye:2007,Blan:Davi:spat:2011}].
The Smith model was excluded because event realizations
from it are too regular to be realistic for practical usage.
Let $h$ be the trivariate difference vector of
longitude, latitude and elevation between two sites.
This vector is transformed into the climate space by $V h$ with
\[
V = \pmatrix {\cos\varphi& \sin\varphi& 0
\vspace*{2pt}\cr
-\sin\varphi/r & \cos\varphi/r & 0
\vspace*{2pt}\cr
0 & 0 & q }, \qquad  r \in(0,1),   \varphi
\in[- \pi/ 2, \pi/2),   q \geq0,
\]
where $\varphi$ is a rotation angle measured counterclockwise
from the east direction, $r$ is the ratio of the minor axis to
the major axis of the ellipse of the geometric anisotropy, and
$q$ gives a weight to elevation in the squared climate distance.
The distance in the climate space is $\sqrt{h^{\top} V^{\top} V h}$,
which is then used in the correlation function of the models.
For comparison, we also fitted isotropic and geometric anisotropic
models in the two-dimensional space without the climate space,
that is, $\varphi= 0$, $r = 1$, and $q = 0$.
Four correlation functions were considered: exponential,
double exponential (also known as Gaussian), Cauchy and
Whittle--Mat\'ern [\citet{Bane:Carl:Gelf:hier:2003}, Section~2.1].
For the Cauchy and Whittle--Mat\'ern correlation, the shape
parameter was fixed at 1 since it is difficult to estimate.
In the geometric Gaussian model, the variation parameter $\delta^2$
which controls the upper bound of the extremal coefficient function
is not easily identifiable jointly with the range parameter
in the correlation function [\citet{Davi:Pado:Riba:stat:2012}].
We fixed $\delta^2$ at $9$ as a compromise between reliable
simulation needed for risk analysis and the near-independence
in pairwise extremal coefficient it can provide.

In total, 24 models were fitted and compared with their CLIC
value conditioning on the marginal GEV models from the first step.
Our final model with the lowest conditional CLIC value ($262280.4$)
is an isotropic geometric Gaussian model which has an
exponential correlation function without elevation effect.
The range parameter is estimated as 4.95, with standard error 0.73.
The spatial dependence decays quickly with distance.
The fitted bivariate extremal coefficient for two sites reaches
1.3 and 1.7 when their distance becomes 19.3 and 149.7 kilometers, respectively.
For illustration, with downtown San Francisco as the reference
point, the extremal coefficients are 1.51, 1.57, 1.89 and 1.92,
respectively, at San Jose, Santa Cruz, Santa Barbara and Los Angles.
The spatial dependence is quite weak, giving much room for the
two-step approach to improve efficiency compared to the pairwise
likelihood approach as shown in the next subsection.

To check the adequacy of the geometric Gaussian model, we first
compared the madogram-based pairwise extremal coefficients
[\citet{Cool:Nave:Ponc:vari:2006}] with those predicted from the model.
The madogram-based pairwise extremal coefficients are calculated
based on the data in the unit Fr\'echet scale obtained
from step~1, instead of ranks, and, hence, it is possible that
some of the estimates exceed the theoretical upper limit 2.
They are plotted against distance in the supplementary material
[\citet{Shan:Yan:Zhan:supp:2014}].
The madogram-based estimators with 100 bins are also shown.
The fitted extremal coefficient curve crosses the scatters
in the middle, suggesting no obvious lack of fit for pairs.
To check the fit beyond pairs, we compared the empirical quantiles
of the maxima of subsets of sites with the quantiles implied from
the model [\citet{Blan:Davi:spat:2011,Davi:Ghol:geos:2012}].
For a subset $A$ of all sites, let $Z_A = \max_{d \in A}Z_d$.
We have observations of $Z_A$ for $n$ independent years\vadjust{\goodbreak}
denoted by $z_{A,1}, \ldots, z_{A,n}$ with $n = 55$.
The distribution of these empirical quantiles can be approximated
from a large number $k$ of simulated realizations from the fitted
model, $z^{\ast}_{A,1,k}, \ldots, z^{\ast}_{A,n,k}$, $ k= 1, \ldots, K$.
The empirical quantiles versus the model-based quantiles for four
subsets of the sites formed geographically based on their latitudes
are plotted in the supplementary material [\citet{Shan:Yan:Zhan:supp:2014}].
The pointwise confidence intervals and simultaneous confidence
bands were obtained from $K = 5000$ simulated realizations
[\citet{Davi:Hink:boot:1997}, Section~4.2.4].
No alarming disagreement between the empirical quantiles and
the model quantiles is observed for any of the subset of sites.

%s6.3 #&#
\subsection{Risk analysis}

For comparison, we also used the pairwise likelihood approach
($M_1$) based on block maxima only to fit the same model as
selected from the two-step approach ($M_2$) with daily records.
Unlike the two-step approach where the site-specific marginal
parameters are estimated separately for each site, the pairwise
likelihood approach needs to estimate all the parameters altogether.
Our profile method updated the marginal parameter estimates
one site at a time first and then updated the dependence parameter;
this process was repeated until convergence.

The point estimates from the two approaches are reasonably close.
For the marginal GEV models, the standard errors from $M_2$ are much
smaller than those from $M_1$ for most of the parameter estimates.
The box plots of the ratio of the standard errors of the
four parameter estimates across 35 sites are presented in
the supplementary material [\citet{Shan:Yan:Zhan:supp:2014}].
In particular, the three quantiles of the ratio are
0.51, 0.58 and 0.65 for the SOI coefficient $\beta_{\mu, s, 1}$, and
0.48, 0.53 and 0.61 for the shape parameter $\beta_{\xi, s, 0}$.
The reduction in standard errors in estimating $\beta_{\mu, s, 1}$
leads to
increased power in detecting the SOI effect: significance
at 5\% was found only at 14 out of 35 sites with $M_1$
(compared to 22 with $M_2$).
The reduction of standard error in estimating $\beta_{\xi, s, 0}$ has important
implications on the accuracy of return level estimation given that
the shape parameter controls the shape of a GEV distribution.
As will be shown next, the reduced standard errors in marginal
parameters lead to more efficient inference about marginal
risk measures such as return levels at each individual site.
For the dependence model, the range parameter was estimated as
6.31 with standard error 0.99 from~$M_1$, in comparison to
4.95 with standard error 0.73 from $M_2$.
The reduction in the standard error in the dependence parameter
estimate of $M_2$
might be explained by its more efficient marginal parameter estimates.

In the spatial context, it is of more interest to see how
the efficiency gain in both marginal and dependence parameter
estimation affects risk measures of jointly defined events.
We first look at the joint 50-year return level for
two sites, as defined in Section~\ref{sec:simu}.
Since SOI is a season-specific covariate, we fix the SOI value
at $-1$, $-0.15$ (the sample average) and $1$ so that the return
levels are interpreted for years with these SOI values separately.
For illustration, consider the three stations near the
Sacramento area: Napa (122.25$^{\circ}$W, 38.27$^{\circ}$N),
Winters (121.97$^{\circ}$W, 38.52$^{\circ}$N) and Davis
(121.78$^{\circ}$W, 38.53$^{\circ}$N); see Figure~\ref{fig:CAelevmap}.
We generated $N = 5000$ realizations of
the model parameters from the approximate multivariate normal
distribution of the estimator from both $M_1$ and $M_2$.
For each realized parameter vector, the joint 50-year return
level was obtained numerically for each pair of the three sites.
Table~\ref{tab:jointrl} shows the 95\% confidence intervals of the
joint 50-year return levels for the three pairs with the empirical
distribution from both $M_1$ and $M_2$ at the three SOI values.
The decreasing trend of the joint return levels as the SOI
value increases is consistent with existing findings
[\citet{Zhan:Wang:Zwie:Groi:infl:2010,Shan:Yan:Zhan:El:2011}].
Interestingly, the confidence intervals from $M_2$ are almost
inside those from $M_1$ for all three pairs, with a reduction
of 27.3\% to 50.7\% in length.

%t2 #&#
\begin{table}
\caption{Joint 50-year return levels (cm) for three pairs at three
different SOI values based on both pairwise likelihood approach
and two-step approach}
\label{tab:jointrl}
\begin{tabular*}{\textwidth}{@{\extracolsep{\fill}}lcccc@{}}
\hline
& \multicolumn{2}{c}{\textbf{Pairwise likelihood (}$\bolds{M_1}$\textbf{)}} & \multicolumn{2}{c@{}}{\textbf{Two-step (}$\bolds{M_2}$\textbf{)}}\\[-6pt]
& \multicolumn{2}{c}{\hrulefill} & \multicolumn{2}{c@{}}{\hrulefill}\\
% \cmidrule(lr){2-3}\cmidrule(lr){4-5}
\textbf{Pair} & \textbf{95\% CI} & \textbf{Width} & \textbf{95\% CI} & \textbf{Width} \\
\hline
& \multicolumn{4}{c}{$\mathrm{SOI} = -1$} \\
Napa \& Winters & (10.22, 15.04) & 4.82 & (10.15, 13.60) & 3.46 \\
Napa \& Davis & (8.62, 12.35) & 3.73 & (8.52, 10.37) & 1.84 \\
Winters \& Davis & (8.33, 11.42) & 3.09 & (8.17, 9.84) & 1.67 \\[3pt]
& \multicolumn{4}{c}{$\mathrm{SOI} = -0.15$} \\
Napa \& Winters & (9.92, 14.79) & 4.87 & (9.85, 13.34) & 3.49 \\
Napa \& Davis & (8.18, 11.87) & 3.68 & (8.30, 10.23) & 1.93 \\
Winters \& Davis & (7.85, 10.83) & 2.98 & (7.90, 9.59) & 1.69 \\[3pt]
& \multicolumn{4}{c}{$\mathrm{SOI} = 1$} \\
Napa \& Winters & (9.34, 14.38) & 5.04 & (9.45, 13.12) & 3.66 \\
Napa \& Davis & (7.36, 11.14) & 3.77 & (7.89, 9.92) & 2.03 \\
Winters \& Davis & (7.05, 10.14) & 3.09 & (7.48, 9.37) & 1.89 \\
\hline
\end{tabular*}
\end{table}

\begin{figure}

\includegraphics{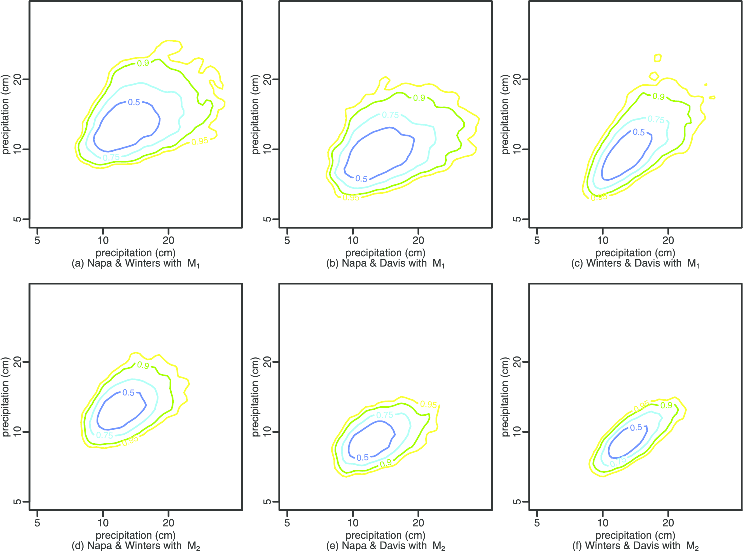}

\caption{Contours of the 50-year sample return levels (cm) for three
pairs on the log scale.
Upper \textup{(a)}, \textup{(b)}, \textup{(c)}: pairwise likelihood approach using block maxima data
($M_1$); Lower \textup{(d)}, \textup{(e)}, \textup{(f)}:
two-step approach ($M_2$). Left \textup{(a)}, \textup{(d)}: Napa~\& Winters;
Center \textup{(b)}, \textup{(e)}: Napa~\& Davis; Right \textup{(c)},
\textup{(f)}: Winters~\& Davis.}
\label{fig:samplerlmaps}
\end{figure}

To gain further insights about the efficiency gain in assessing
bivariate risk measures, we investigated the joint sampling
distribution of the site-wise maximum extremal precipitations
over every 50 years for all pairs of the three sites.
Realizations from the distribution can be drawn for the three
sites and then used to assess their joint behavior.
The SOI was fixed at the sample average $-0.15$ for ease of interpretation.
For each of the $N = 5000$ parameter vectors drawn
from their asymptotic normal distribution, we generated
50 years of data and obtained the sitewise maxima.
On the log scale, Figure~\ref{fig:samplerlmaps} shows the
empirical contours of the 5000 draws from the sampling
distribution for the three sites with both $M_1$ and $M_2$.
The levels with 50, 75, 90 and 95 percent of coverage are plotted.
It is apparent that the joint sampling distribution is
much more compact from $M_2$ than from~$M_1$.
Consequently, much tighter approximate confidence regions are
obtained with $M_2$ than with $M_1$.
Positive dependence between each pair is clearly visible,
with especially stronger dependence between the last pair
(Winters and Davis), which is explained by their distance.

%f3 #&#

%s7 #&#
\section{Discussion}
\label{sec:disc}

In contrast to the pairwise likelihood approach which utilizes
only block maxima, the two-step approach uses more information through
daily records and makes more efficient inferences about the parameters.
The consistency in marginal GEV parameter estimation is not
affected by possible misspecification of the dependence model.
Our simulation study showed appreciable efficiency gain of the
two-step approach in comparison to the pairwise likelihood approach.
The two-step approach is simple to implement with existing software,
intuitive for practitioners, and avoids defining multivariate thresholds
[\citet{Wads:Tawn:depe:2012,Bacr:Gaet:esti:2014,Huse:Davi:spac:2014}]
or\vadjust{\goodbreak} multivariate Pareto process modeling [\citet{Aulb:Falk:test:2012}].
A caveat is that, as with the POT approach or the point process approach,
not only the block maxima but also the exceedances over the threshold
need to have distributions that are well approximated by the corresponding
limiting distribution; not meeting the requirement may lead to bias
as illustrated in our simulation study.

In application to maximum daily winter precipitation in California,
large scale climate variation ENSO was found to have significant
negative impact on the location parameter of the marginal
GEV distribution at 22 out of 35 sites with the two-step approach
(compared to 14 with the pairwise likelihood approach).
Risk analysis with the two-step approach gives much tighter
confidence intervals and confidence regions for joint risk measures
than the pairwise likelihood approach.

Several methodological aspects merit further investigation.
In the first step, we did not address threshold selection, an
important and still active problem even for univariate extreme value
analysis [e.g., \citet{Guil:Hall:diag:2001,Thom:etal:auto:2009}].
Recent research has shown a promising approach with quantile regression for
nonstationarity with covariate information [\citet{Nort:Jona:thre:2011}].
Alternatively, one may use the $r$ largest order statistic to
construct the marginal likelihood [e.g., \citet{Cole:an:2001}].
Compared to the bivariate threshold-based approaches, the two-step
approach may potentially be less efficient if the distributional
approximation over the bivariate threshold is accurate, but its
marginal inference is robust to dependence structure misspecification.
A study on the robustness-efficiency trade-off would be interesting.

\section*{Acknowledgments}
We thank the reviewers and the Associate Editor for their constructive comments.

\begin{supplement}[id=suppA]
\stitle{Additional simulation results and data analysis\\}
\slink[doi]{10.1214/14-AOAS804SUPP} %[doi,text={...}] - jei reikia suskaldyti doi
\sdatatype{.pdf}
\sfilename{aoas804\_supp.pdf}
\sdescription{We provide a sandwich variance estimator,
additional tables summarizing the simulation study and
additional figures in analyzing the California precipitation data.}
\end{supplement}

% imsref loaded by akundreckaite, 2015-01-27 08:14:58
% imsref loaded by akundreckaite, 2015-01-27 08:34:10
%
% imsref loaded by akundreckaite, 2015-01-30 13:46:47


%suskaldyti doi

\printaddresses
\end{document}